\documentclass[10pt,twocolumn,letterpaper]{article}

\usepackage{iccv}
\usepackage{times}
\usepackage{graphicx}
\usepackage{amsmath}
\usepackage{amssymb}
\usepackage{textcomp}
\usepackage{float}
\usepackage{adjustbox}
\usepackage{subfigure}
\usepackage{textcomp}
\usepackage{csquotes}

\date{}

\iccvfinalcopy %

\def\iccvPaperID{1908} %

\ificcvfinal\fi

\begin{document}
\title{Feature-Align Network with Knowledge Distillation for Efficient Denoising}

\author{
\begin{tabular}{c@{\hspace{0.4in}}c@{\hspace{0.4in}}c}
Lucas D. Young$\thanks{Equal contribution.}$ & Fitsum A. Reda$^*\thanks{Affiliated with Facebook at the time of this work.}$ & Rakesh Ranjan \\
Jon Morton &  Jun Hu &  Yazhu Ling \\
Xiaoyu Xiang & David Liu & Vikas Chandra 
\end{tabular} \\
Facebook Inc.
}

\maketitle
\ificcvfinal\thispagestyle{empty}\fi

\begin{abstract}
We propose an efficient neural network for RAW image denoising. Although neural network-based denoising has been extensively studied for image restoration, little attention has been given to efficient denoising for compute limited and power sensitive devices, such as smartphones and smartwatches. In this paper, we present a novel architecture and a suite of training techniques for high quality denoising in mobile devices. Our work is distinguished by three main contributions. (1) Feature-Align layer that modulates the activations of an encoder-decoder architecture with the input noisy images. The auto modulation layer enforces attention to spatially varying noise that tend to be "washed away" by successive application of convolutions and non-linearity. (2) A novel Feature Matching Loss that allows knowledge distillation from large denoising networks in the form of a perceptual content loss. (3) Empirical analysis of our efficient model trained to specialize on different noise subranges. This opens additional avenue for model size reduction by sacrificing memory for compute. Extensive experimental validation shows that our efficient model produces high quality denoising results that compete with state-of-the-art large networks, while using significantly fewer parameters and MACs. On the Darmstadt Noise Dataset benchmark, we achieve a PSNR of 48.28dB, while using 263$\times$ fewer MACs, and 17.6$\times$ fewer parameters than the state-of-the-art network, which achieves 49.12dB.

\newcommand{\teaserresultswidth}{0.472\linewidth}
\begin{figure}[t]
\begin{center}
  \subfigure[Noisy Input,  PSNR = 18.76]{
  \includegraphics[width=\teaserresultswidth]{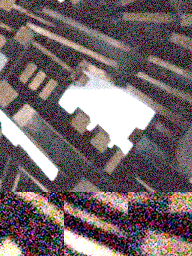}
  \label{subfig:teaser2}
  }
  \subfigure[Ground Truth]{
  \includegraphics[width=\teaserresultswidth]{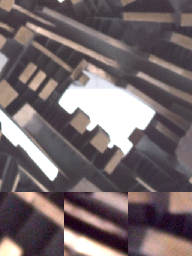}
  \label{subfig:teaser1}
  }
  \subfigure[CycleISP~\cite{cycleisp}, PSNR = 40.44, 702.73 GMACs/MP]{
  \includegraphics[width=\teaserresultswidth]{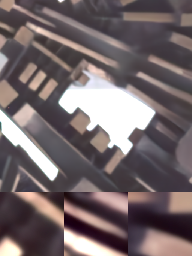}
  \label{subfig:teaser3}
  }
  \subfigure[Our Model, PSNR = 36.33, 2.67 GMACs/MP]{
  \includegraphics[width=\teaserresultswidth]{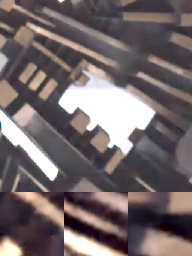}
  \label{subfig:teaser4}
  }
\end{center}
  \caption{Our efficient denoising method compared to the existing state-of-the-art denoising method, CycleISP when applied to an image in the Darmstadt Noise Dataset ~\cite{darmstadt}. Our method produces similar results with 263x fewer MACs.}
\label{fig:teaser}
\end{figure}

\end{abstract}

\section{Introduction}

Image acquisition is inevitably contaminated by noise due to various environmental effects. Noisy images are especially more prevalent in small devices such as smartphones and wearables. Often these devices are characterized by small sensors and limited light intake ability, leading to low perceptual quality images. Thus, efficient image denoising algorithms are highly desirable to restore image quality in mobile devices.

Image denoising is a classical yet actively studied topic in image restoration~\cite{bm3d, nlm, sparse3d, classic2, classic3, classic4}. Recent attention for the problem focuses on applying deep neural networks to RAW sensor data or to images obtained after post-processing with the device's image signal processor (ISP)~\cite{brooks2019unprocessing, gms, pseudoisp, mirnet, dualpath, pridnet, n3net, cycleisp}. 
Most of these networks are based on the U-Net~\cite{U-Net} architecture. They produce high quality denoising and leverage large quantities of training data. They are, however, computationally too expensive to run at real-time on compute and power sensitive edge devices. 

The success of the U-Net architecture~\cite{U-Net} has been celebrated in learning complex mappings for several dense prediction tasks, such as semantic segmentation~\cite{WangSCJDZLMTWLX19}, depth estimation~\cite{luo2020consistent} or image synthesis~\cite{wang2018high}. However, its direct deployment on mobile devices with limited compute-budget is not optimal. Trivial reduction of U-Net parameters is not sufficient to learn robust models. As such, efforts have been spent on efficient building blocks. MobileNets~\cite{howard2017mobilenets, singh2019shunt, howard2019searching} proposed a suite of efficient blocks for efficient learning, starting with a depth-wise convolution, inverted residuals, and later with optimized blocks searched via neural architecture search algorithms. Similarly, \cite{hu2018squeeze} introduced squeeze-excitation-networks, and \cite{zhang2018shufflenet} proposed Shuffle Nets that proved to be effective for mobile applications.

Another recent phenomenon observed in U-Net is that successive application of convolutions and non-linearity tend to \enquote{wash away} important image cues in deeper layers. This phenomenon was first studied in \cite{park2019semantic} for the task of conditional image synthesis and shown to lead to sub-optimal image synthesis outcomes. The same work addresses this issue with a spatially-adaptive normalization layer. Specifically, it modulates the U-Net features with transformations learned from the input semantic layout, and shows improved learning outcomes in image synthesis. In our experiments, we observed this phenomenon to be even more pronounced in light-weight image denoising U-Net architectures.

In this work, we build upon the U-Net~\cite{U-Net} architecture and introduce an efficient neural network for high quality RAW image denoising. We employ a variant of the MobileNetV2~\cite{singh2019shunt} efficient module in place of normal convolutions. Specifically, our variant MobileNetV2~\cite{singh2019shunt} uses group convolutions in place of depth convolutions, which further improves memory efficiency. To alleviate the \enquote{washing away} of features in U-Net, we take inspiration from the SPADE architecture~\cite{park2019semantic} and propose to modulate the convolutional features at each layer with the input noisy images. This auto modulation of features with transformations learned directly from the input noisy images allows our models to attend to spatially varying noise. Experimental validations suggest the effectiveness of this mechanism in our efficient U-Net architecture. We refer to our modulation layer as \enquote{Feature-Align}.

Student-teacher learning has proven to be effective in learning efficient models~\cite{wang2021knowledge}. Knowledge distillation allows the transfer of richer knowledge to light-weight models. Our experiments also prove the effectiveness of knowledge distillation in image denoising, especially in settings with limited training datasets. In this work, we build up on the student-teacher training mechanism and propose a new Feature Matching Loss that performs knowledge distillation in the form of a perceptual content loss. Specifically, we extract deep multi-scale features from our efficient model's output image and optimize them to match the features extracted from the clean ground-truth image, in a similar way as the VGG perceptual loss~\cite{johnson2016perceptual}. One distinction about our features matching loss is that we use a large pre-trained image denoising network, for instance ~\cite{brooks2019unprocessing}, to perform the feature extraction, instead of ImageNet~\cite{deng2009imagenet} trained image classification networks~\cite{simonyan2014very}. Using a domain-specific feature extractor, in our case an image denoiser, allows the transfer of knowledge at deeper representation levels. As we will show in experiments, our Feature Matching Loss leads to more crisp and realistically denoised images than that of training with standard student-teacher distillation or regular perceptual losses. 

Our work also explores additional avenues to model size reduction. Specifically, we train even smaller models targeting for specific noise subranges. Use of an array of light-weight specialized models allows us improve inference time at the the expense of memory in mobile devices. Experimental validations suggest the effectiveness of this technique.

In summary, the primary contributions of this paper are:
\begin{itemize}
    \item Feature-Align layer that modulates the activations of an encoder-decoder architecture with the input noisy images. The auto modulation layer enforces attention to spatially varying noise that tend to be \enquote{washed away} by successive application of convolutions and non-linearity.
    \item A  novel  feature  matching  loss  that allows knowledge distillation from large denoising networks in the form of a perceptual content loss.
    \item Empirical analysis of our efficient model trained to specialize on different noise subranges, which opens additional avenue for model size reduction by sacrificing memory for compute.
\end{itemize}

\section{Related Work}

Single-frame image denoising is a fundamental problem in image processing and computer vision. Prior methods such as BM3D~\cite{bm3d} and non-local~means~\cite{nlm} rely on hand-engineered algorithms~\cite{sparse3d, classic2, classic3, classic4}. With the introduction of data-driven neural network-based methods~\cite{gms, pseudoisp, mirnet, dualpath, pridnet, n3net}, datasets with paired noisy and noise-free images became sought after~\cite{sidd, sid, renoir}. However, it is well known and demonstrated in public benchmarks that performing denoising in the RAW domain yields superior results than denoising on the final RGB images~\cite{darmstadt}. To help with the scarcity of paired noisy and clean RAW images, other works proposed to simulate noisy images. Clean images are post-processes by a simple addition of white gaussian noise (AWGN) to simulate noisy images. Recent works observe that true sensor noise exhibits characteristics unlike AWGN, and thus they propose robust noise modeling mathematics by considering the physical properties of sensors ~\cite{brooks2019unprocessing, darmstadt, practical, azzari, hasinoff_capture, hasinoff_textbook, foi}. The work of~\cite{brooks2019unprocessing} and CycleISP~\cite{cycleisp} further synthesize RAW images from real-life RGB images by inverting the ISP pipeline. Public real-life image datasets, such as MIRFLICKR~\cite{mirflickr}, are leveraged with artificial noise modeling to scale-up the training dataset. This approach has been demonstrated in achieving state-of-the-art performance on public benchmarks, suggesting the generalizability of the approach.

While these works introduce valuable techniques for denoising, they do not study less computationally expensive variants of their methods. PMRID~\cite{practical} is the first to address this disparity in neural denoising. Specifically, it introduces an efficient model architecture using separable convolutions. It also builds upon the variance stabilizing transform~\cite{vst1, vst2} by introducing the k-sigma transform, which normalizes data such that per-pixel variance is not dependent on the ISO of the exposure. PMRID also describes a noise modeling method similar to~\cite{brooks2019unprocessing}. The work we propose here shares similar goals to PMRID. However, a direct quantitative comparison could not be performed at the time of this paper's writing because PMRID's trained model or test datasets are not publicly available.

Our proposed technique builds upon previous works and introduces an efficient denoising neural network and a suite of training techniques. In particular, it incorporates the following prior techniques:
\begin{itemize}
\item Noise modeling of our target camera's sensor, Sony IMX258, with a method similar to PMRID and UPI, to add realistic artificial noise to ground truths.
\item Unprocessing~\cite{brooks2019unprocessing} the MIRFLICKR-1m ~\cite{mirflickr} dataset to create a large RAW training dataset.
\item Incorporation of Bayer Augmentation~\cite{bayeraug} in our training pipeline.
\item Use of Bayer Unification~\cite{bayeraug}
to use ground truths from the sensors in the SIDD~\cite{sidd} and SID~\cite{sid} dataset.
\item Incorporation of the k-sigma transform~\cite{practical}
\end{itemize}

Lastly, we combine inspiration from a method traditionally applied to super-resolution, perceptual loss~\cite{johnson2016perceptual}, and knowledge distillation commonly applied for student-teacher knowledge distillation to propose a novel Feature Matching Loss function. As we will show in experiments, our new loss produces results that are better than using either of these established techniques.

\section{Method}
Given a noisy RAW image $\mathbf{I}_{n}$, we aim to learn efficient neural networks to denoise and produce clean RAW image $\widetilde{\mathbf{I}}$, in a data-driven way. Mathematically, the denoising problem is formulated as, 

\begin{equation}
    \widetilde{\mathbf{I}} = \mathcal{F}(\mathbf{I}_{n}), 
\end{equation}
where $\mathcal{F}$ is an efficient neural network we aim to learn. We adopt the techniques introduces in~\cite{brooks2019unprocessing} to create synthetic RAW training pairs. Denote the noise sampled from a noise model $\mathbf{n}$ and the the unprocessed \enquote{clean} RAW image $\mathbf{I}$. 
Note that, the RGB images we present here to visually motivate or validate our methods are obtained after processing the denoised RAW image $\widetilde{\mathbf{I}}$ with the camera ISP. We use ${\mathbf{I}}_{n}^{f}$ and $\widetilde{\mathbf{I}}^{f}$ to denote the final noisy and final denoised RGB image, respectively. 

\subsection{Model Architecture}
\begin{figure*}[htb!]
\centering
\includegraphics[width=1.9\columnwidth]{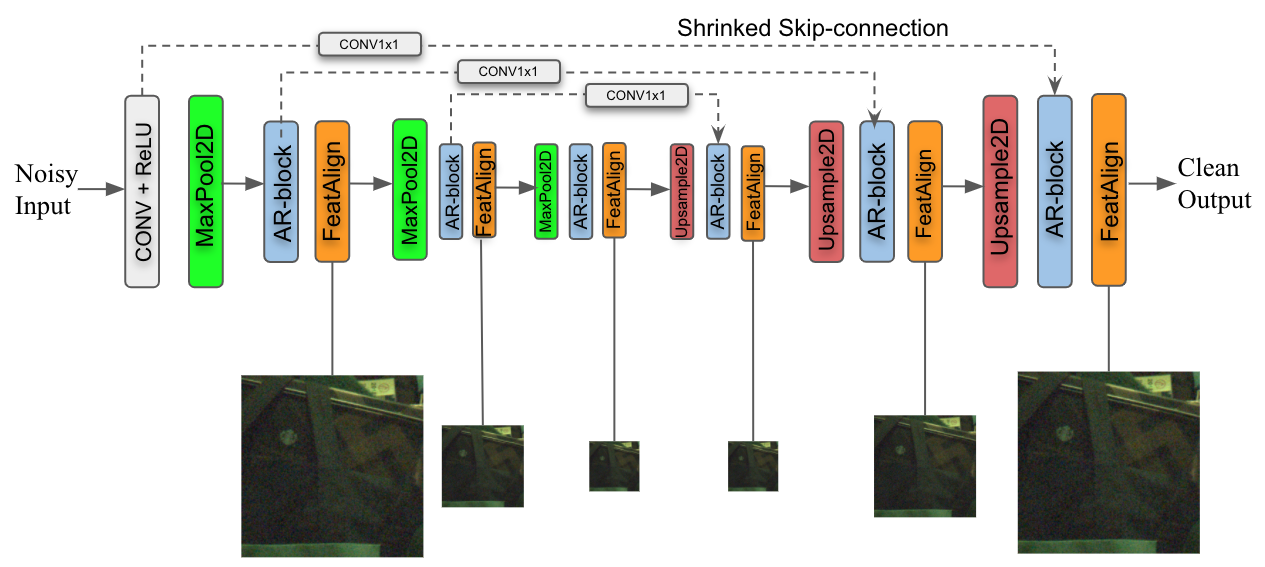}
\caption{Our U-Net-based model architecture. Shrinked skip-connections combine high-resolution features in encoding layers with low-resolution features in decoder layers. A Feature-Align layer combines the original input features at different resolutions with the output for each efficient ARNet block.}
\label{fig:denoising_arch}
\end{figure*}

We realize our neural network $\mathcal{F}$ with an efficient U-Net~\cite{U-Net} architecture.
We propose techniques to boost the learning ability light-weight U-Net models. In particular, we make the following architectural changes: 1) To achieve efficient learning, we replace the convolutional layers with a variant of the MobileNet-V2~block~\cite{sandler2018mobilenetv2}. 2) To reduce the memory footprint of skip layers, we introduce shrinked skip-connection, implemented with a point-wise convolution. 3) To minimize the "washing away" phenomenon of important image features in deep convolutional layers, as pointed out in the SPADE architecture~\cite{park2019semantic}, we introduce a new Feature-Align layer. The following subsections describe each design choice. Figure \ref{fig:denoising_arch} illustrates our proposed efficient architecture.

\begin{figure}[ht!]
\centering
\includegraphics[width=0.8\columnwidth]{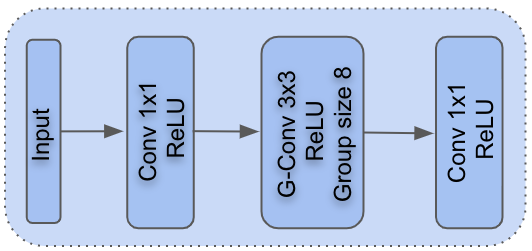}
\caption{ARNet-block, a variant of MobileNet-V2~\cite{sandler2018mobilenetv2}}
\label{fig:arnet_block}
\end{figure}

\noindent \textbf{A variant of MobileNet-V2 block:} 
In this work, we use a variant of the MobileNet-V2 efficient block to realize each contracting and expanding layer in our U-Net. Specifically, as illustrated in Figure \ref{fig:arnet_block}, we replace MobileNet-V2's depth-wise convolution with a group convolution. This choice allows efficient memory access.

\noindent \textbf{Shrinked Skip Connections:} Skip-connections in a U-Net architecture allow joint processing of multi-scale features. High-resolution features from the contracting layers are combined with low-resolution features from the expanding layers in the decoder layer. In memory constrained devices, storing encoder activations until their processing by decoder layers could be memory inefficient. To alleviate this, as illustrated in Figure \ref{fig:denoising_arch}, we apply a point-wise convolution to shrink the encoder features. At the decoder layer, we replicate the contracted features to match the size of the expanding features to fusion the features. 
 
\noindent \textbf{Input Feature-Align Layer:} Prior neural network-based denoising techniques take in noisy inputs and directly process them with a stack of convolutions and non-linearity. While this is also a common technique for various dense prediction tasks, in our experiments, we observed such direct processing to bias our light-weight models to exhibit global over- or under-denoising. As shown in our experiments, we attribute this to the recently studied phenomenon of "washing away" of input details during processing with a stack of multi-scale layers. In our case, we tightly integrate the input noisy information at each layer of our network with a Feature-Align layer, to enable our models to learn to denoise images in a spatially adaptive manner. 
\begin{figure}[htb!]
\centering
\includegraphics[width=0.8\columnwidth]{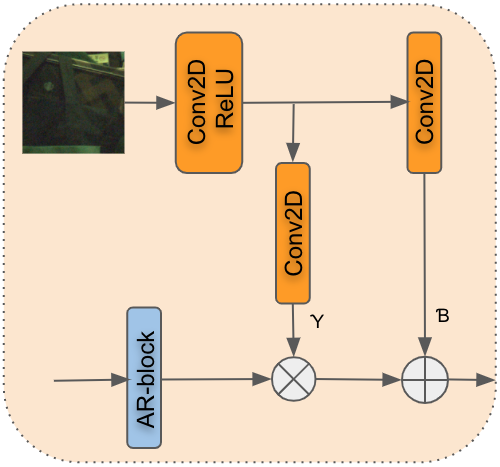}
\caption{Input Feature-Align Layer, inspired by the SPADE architecture~\cite{park2019semantic}}
\label{fig:feature_align}
\end{figure} 

Let $\mathbf{F}^{i} \in \mathbb{R}^{N\times C^{i}\times H^{i}\times W^{i}}$ denote the feature map of layer ${i}$, with $N, C^{i}, H^{i}, W^{i}$ being the batch size, channel count, height and width, respectively, of the feature map. Our Feature-Align layer, in a similar way as a Batch Norm layer, or the Spatially-Adaptive De-normalization (SPADE) layer, applies a scale and a bias to affinely transform the inputs, guided by the noisy input image. Specifically, we compute the pixel-wise scale and bias parameters based on the noisy input image, as illustrated in Figure \ref{fig:feature_align}. Mathematically, the affine transformation of a feature map is given by 
\begin{equation}
    \bold{G}^{i} = \boldsymbol{\gamma}^{i} \cdot \bold{F}^{i} + \boldsymbol{\beta}^{i}, 
\end{equation}
with $\boldsymbol{\gamma}^{i} \in \mathbb{R}^{N\times C^{i}\times H^{i}\times W^{i}}$ and $\boldsymbol{\beta}^{i} \in \mathbb{R}^{N\times C^{i}\times H^{i}\times W^{i}}$ being the scale and bias parameters that are a function of the input noisy image.

\subsection{Loss functions}

\noindent \textbf{Charbonnier Loss:}
Our primary loss function is the Charbonnier Loss~\cite{charb} over the denoised image, given by 
\begin{equation}
    \mathcal{L}_{charb} = \sqrt{(\widetilde{\mathbf{I}} - \mathbf{I})^2 + c^2}
\end{equation}
where $c$ is a constant parameter ($1e-6$ in our implementation). 

\noindent \textbf{RGB Perceptual Loss:}
UPI~\cite{brooks2019unprocessing} describes a variant of their network in which training loss is computed after postprocessing the predicted image and ground truth with a minimalist ISP implementation. This incentivizes the network to predict RAW images that after postprocessing, yield RGB images that closely match the ground truth after postprocessing, rather than to optimize for RAW signal fidelity. We expand upon this idea with the RGB Perceptual Loss, in which a perceptual loss such as that described in~\cite{johnson2016perceptual} is computed upon the predicted image and ground truth after postprocessing. After the postprocessing step, the implementation is identical to that of ~\cite{johnson2016perceptual}; the loss is defined as the difference between $\widetilde{\mathbf{I}}^{f}$ and $\mathbf{I}^{f}$ in content and style in activations of a VGG16 classification network ~\cite{vgg} pretrained on ImageNet.

\noindent \textbf{Simple Knowledge Distillation:}
The additional layers of conventional larger networks can enable the model to encode image features at a higher, more perceptual level and make imaginative judgments to "fill in the blank" of a missing texture obscured by noise. Since this capability is diminished with smaller networks, image outputs from efficient networks are prone to appear oversmoothed. As a result, knowledge distillation~\cite{hinton2015distilling} is applicable to the challenge of creating an efficient model with perceptually pleasing image outputs.

Another observation from experimenting with a large denoising model (the same U-Net architecture used in ~\cite{brooks2019unprocessing}) is that a denoising network can learn to perform the identity transformation as desired - that is, if a noise-free image is fed to the network as an input, the output is virtually unaltered. This observation enables a simple implementation of knowledge distillation in the form of a loss function defined as the difference between the student network's predicted image, $\widetilde{\mathbf{I}}_{student}$ and the large teacher network's predicted image, $\widetilde{\mathbf{I}}_{teacher}$:

\begin{equation}
    \mathcal{L}_{k.d.} = |\widetilde{\mathbf{I}}_{student} - \widetilde{\mathbf{I}}_{teacher}|
\end{equation}

\noindent \textbf{Feature Matching Loss for Knowledge Distillation:} 
Expanding on Simple Knowledge Distillation and perceptual loss, our novel Feature Matching Loss implements knowledge distillation in the form of a perceptual loss function. The Feature Matching Loss is defined as the difference in content and style between the activations of the teacher network on the predicted image and the ground truth image. The implementation is identical to~\cite{johnson2016perceptual}, except the classifier model is substituted for the teacher network. By combining knowledge distillation with perceptual loss, the capability of large networks to recover fine details obscured by noise is specifically targeted for distillation.

\subsection{Noise Subrange Model Array}
The number of parameters in our efficient model architecture is limited in comparison to large model architectures traditionally used for neural denoising. In contrast, the problem space encountered in photography on mobile devices that typically have cameras with small aperture sizes widely ranges from minimal noise to extreme noise, requiring vastly different denoising strategies depending on the noise level. With a limited number of parameters, it is challenging to create an efficient network that optimally responds to every possible noise level. Thus, we offer the method of a Noise Subrange Model Array (NSMA), in which the range of noise levels is partitioned into $n$ subranges where an individual model is trained for each noise subrange. To partition the noise levels, we refer to the regression between $\log a$ and $\log b$ in Figure \ref{fig:regressions} of the supplemental material and partition the x-axis into $n$ parts. Given the global minimum signal dependent noise parameter, $a_{min}$, and global maximum signal dependent noise parameter $a_{max}$, Equation \ref{eq:nsma1} and \ref{eq:nsma2} describe the minimum and maximum $a$ parameters for artificial noise in training for the zero-indexed $i$th model in the array of $n$ models.

\begin{equation}
    \label{eq:nsma1}
    \log a_{min_{i}} = \log a_{min} + i / n \cdot (\log a_{max} - \log a_{min})
\end{equation}
\begin{equation}
    \label{eq:nsma2}
    \log a_{max_{i}} = \log a_{min} + (i + 1) / n \cdot (\log a_{max} - \log a_{min})
\end{equation}

In testing, we observe the annotated $a$ noise parameter of the image as described in supplemental Section \ref{sec:noise} and select the corresponding model from the model array.

\subsection{Training and Implementation Details}
\label{sec:trainingdetails}
Our models are trained via adding synthetic noise to ground truths sourced from \cite{sidd}, \cite{sid}, and the unprocessing method described in \cite{brooks2019unprocessing}. Bayer Augmentation and Unification~\cite{bayeraug} and the k-sigma transform~\cite{practical} are incorporated into our method. A full description of training and implementation details is provided in supplemental Section \ref{sec:traindetails}

\subsection{Test Dataset}
\label{sec:datasets}

While our denoising network is trained with artificial noise, we test our network with real noise. We contribute the publicly available Feature-Align Paired Test Dataset of noisy RAW images paired with corresponding noise-free ground truths. This dataset consists of carefully aligned pairs of noisy short exposures that correspond to noise-free long exposures. The method used to collect this dataset is provided in supplemental Section \ref{sec:testdetails}.

\section{Results}

\subsection{Experiments}

In this section we evaluate denoising results of various experiments using the paired noisy and ground truth test set described in Section \ref{sec:datasets}. We evaluate our results using traditional metrics for image quality, PSNR and SSIM, both before and after processing the images using the minimalist ISP pipeline from~\cite{brooks2019unprocessing} with modified white balance gains for our sensor. We also evaluate our results qualitatively against the ground truth image. To display texture preservation, all images are zoomed in to the same highly textured region of an image in the test set. We quantify the computational efficiency of networks by the number of multiply-accumulate operations (MACs) required by the network.

\begin{figure}
    \centering
    \adjustimage{width=1.0in,valign=m}{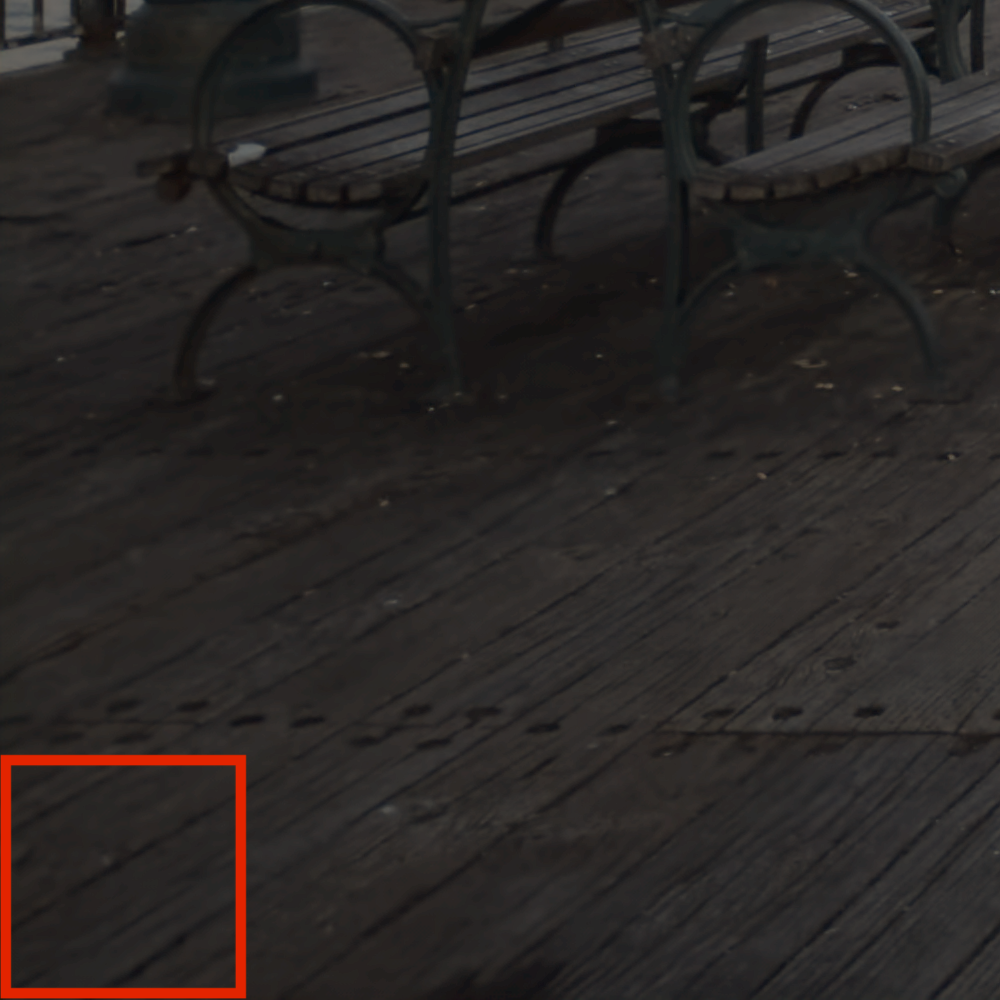}
    \adjustimage{width=1.0in,valign=m}{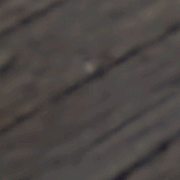} \adjustimage{width=1.0in,valign=m}{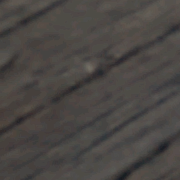}\\
    \adjustimage{width=1.0in,valign=m}{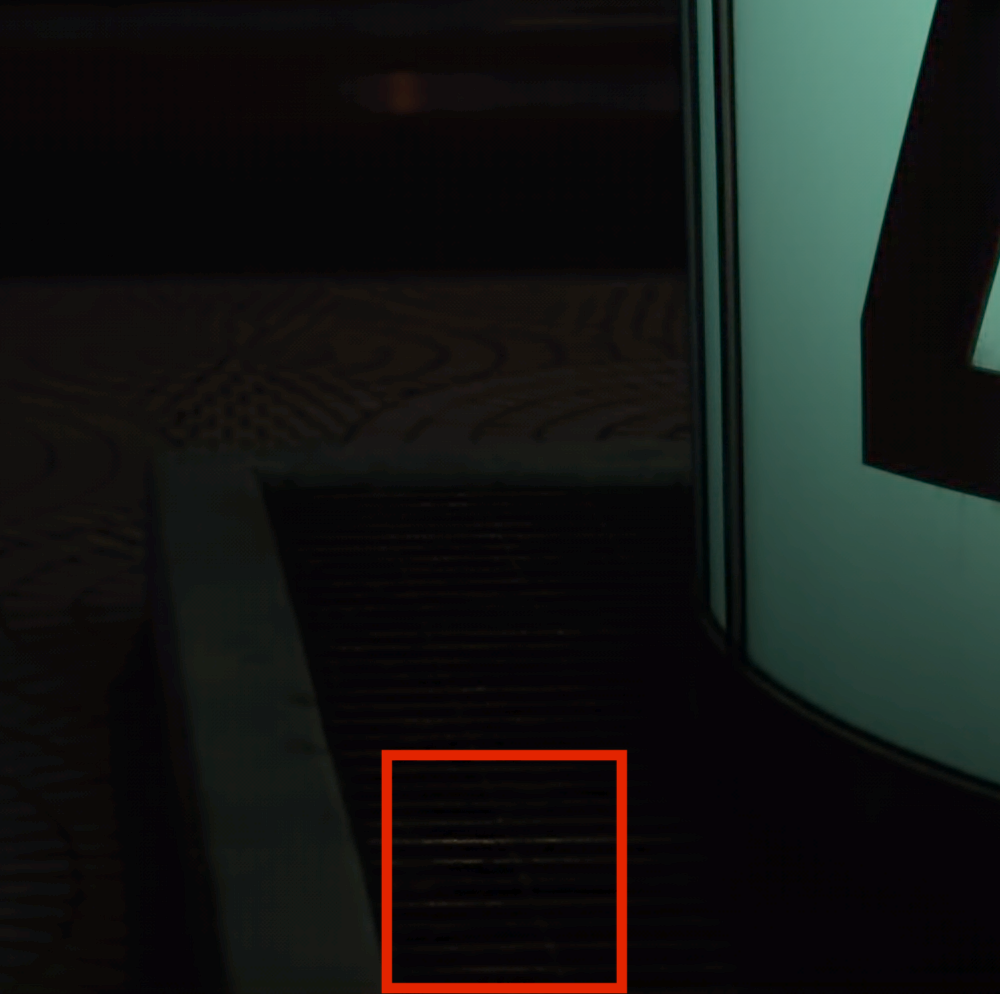}
    \adjustimage{width=1.0in,valign=m}{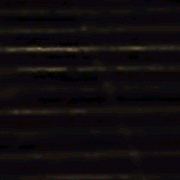} \adjustimage{width=1.0in,valign=m}{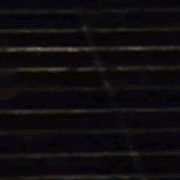}\\
    \adjustimage{width=1.0in,valign=m}{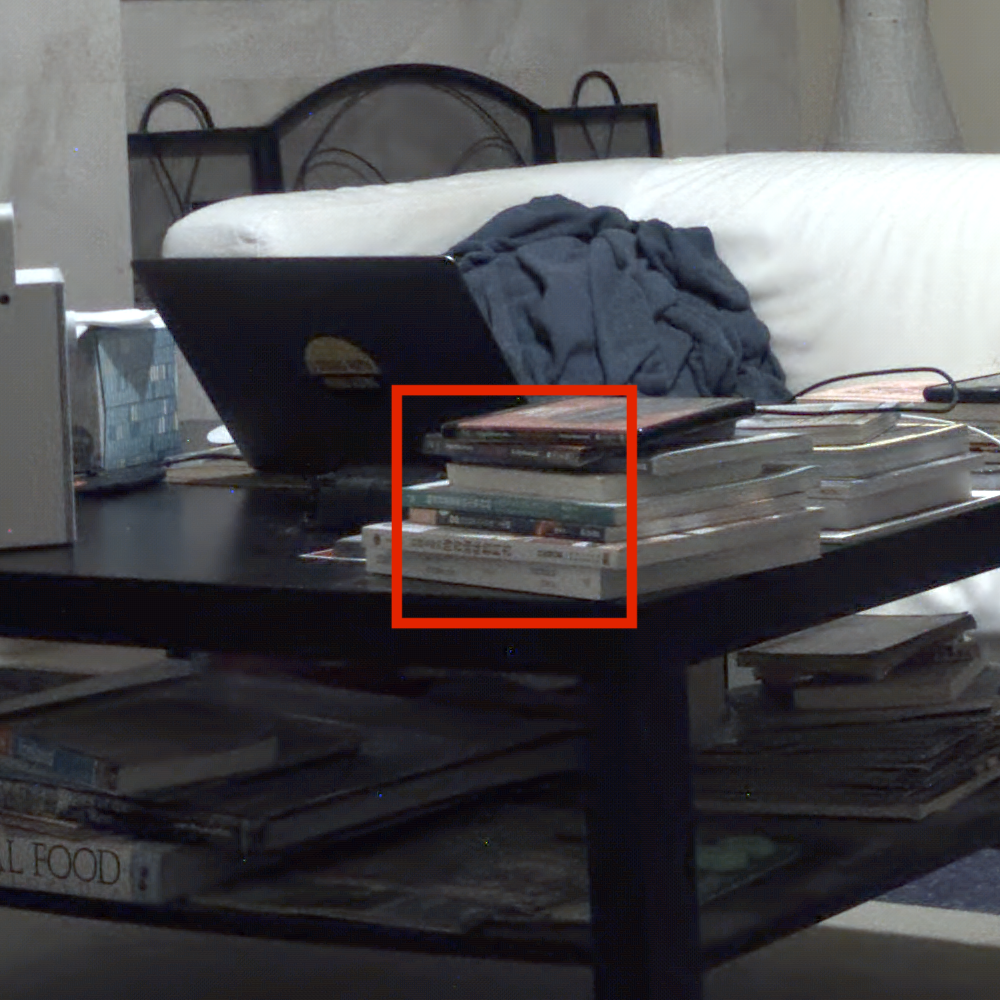}
    \adjustimage{width=1.0in,valign=m}{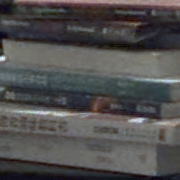} \adjustimage{width=1.0in,valign=m}{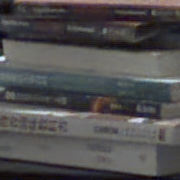}\\
    \caption{Left: Source image from Google Pixel-4. Red box highlights selected crop for comparison. Middle: Model output with wavelet transform. Right: Model output with our Feature-Align layer. Middle and right images are brightened for clarity.}
    \label{fig:spadevswav}
\end{figure}

Table \ref{table:model_architecture} compares our Feature-Align method against a similarly sized model that uses the wavelet transform in upsampling and downsampling as a baseline to approximate the prior state-of-the-art~\cite{waveletdeep, classicwavelet}. To simplify the comparison, the Noise Subrange Model Array is not used in this experiment. A plain U-Net that does not include the wavelet transform or the Feature-Align layer, along with the Feature-Align model with shrinked skip connections removed are also included in the comparison. With traditional metrics, the Feature-Align model performs virtually equivalently to the wavelets model. Qualitatively, the Feature-Align model shows sharper edges and less blurring of fine details. To show additional generalization of the Feature-Align layer, we compare the wavelet transform model to the Feature-Align model on collected noisy RAW Google Pixel-4 inputs in Figure \ref{fig:spadevswav}. We note that the qualitative benefits of the Feature-Align layer are more easily observed on our Pixel-4 inputs than our paired test dataset. Furthermore, the Feature-Align model without shrinked skip connections performs the worst among the 4 networks. This indicates that the Feature-Align layer is not redundant to the shrinked skip connections despite the similarity of these techniques.

In Table \ref{table:nsma} we examine the effect that the NSMA has on the Feature-Align model. The NSMA, which is comprised of 4 models, improves performance in every metric and demonstrates superior recovery of texture.

Using the Feature-Align model and NSMA, in Table \ref{table:loss} we compare the effect of altering the loss function. Since RGB Perceptual Loss, Simple Knowledge Distillation, and Feature Matching Loss are targeted at perceptual image quality rather than pure signal fidelity, for a fair comparison we include the perceptual metric, LPIPS~\cite{lpips}. A lower LPIPS score indicates greater perceptual similarity. Simple Knowledge Distillation performs the best in the traditional PSNR and SSIM metrics, while Feature Matching Loss performs the best in the LPIPS metric. Qualitatively, the outputs from the Feature Matching Loss models are sharper and have the most texture present. Charbonnier Loss outperforms Feature Matching Loss in the traditional metrics, further demonstrating that the traditional metrics tend to favor oversmoothing over texture reconstruction.

\begin{figure}
    \centering
    \adjustimage{width=1.0in,valign=m}{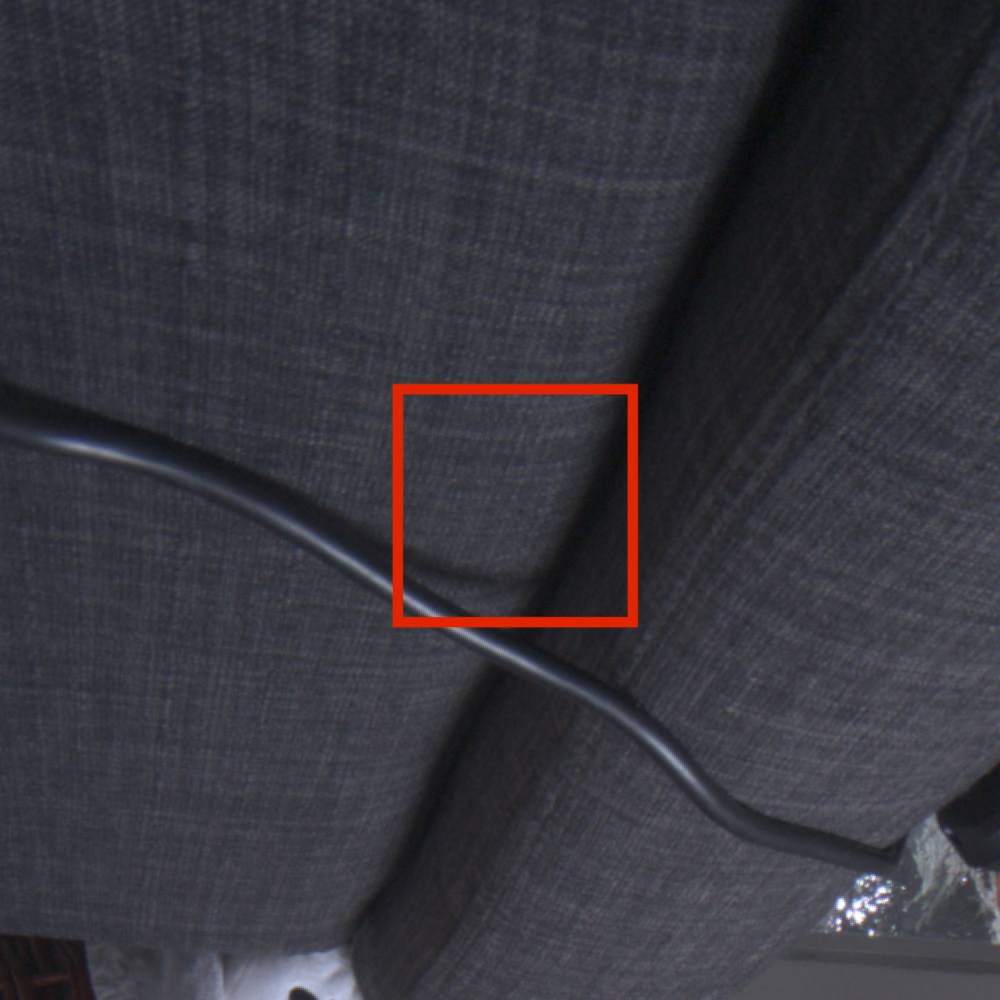}
    \adjustimage{width=1.0in,valign=m}{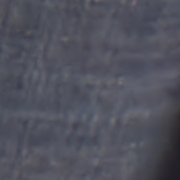} \adjustimage{width=1.0in,valign=m}{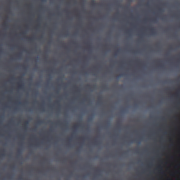}\\
    \adjustimage{width=1.0in,valign=m}{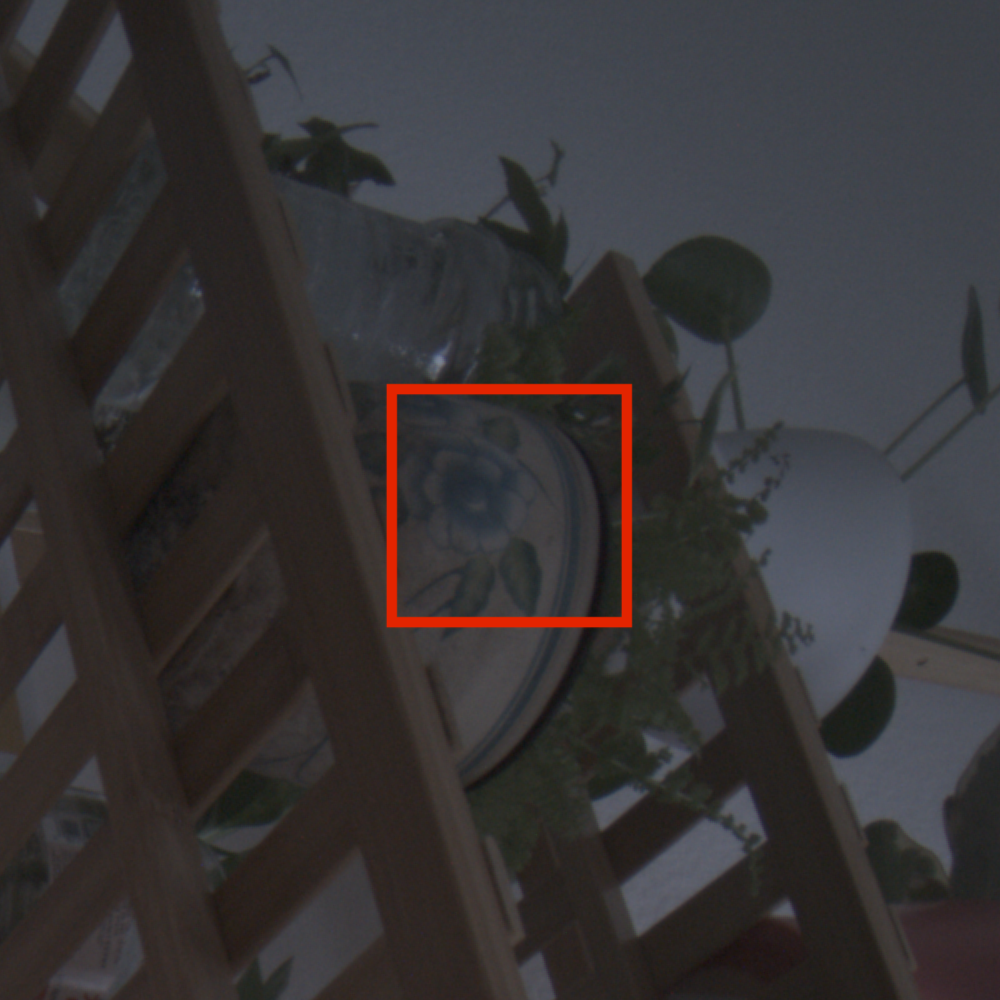}
    \adjustimage{width=1.0in,valign=m}{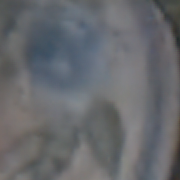} \adjustimage{width=1.0in,valign=m}{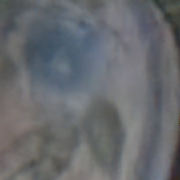}\\
    \adjustimage{width=1.0in,valign=m}{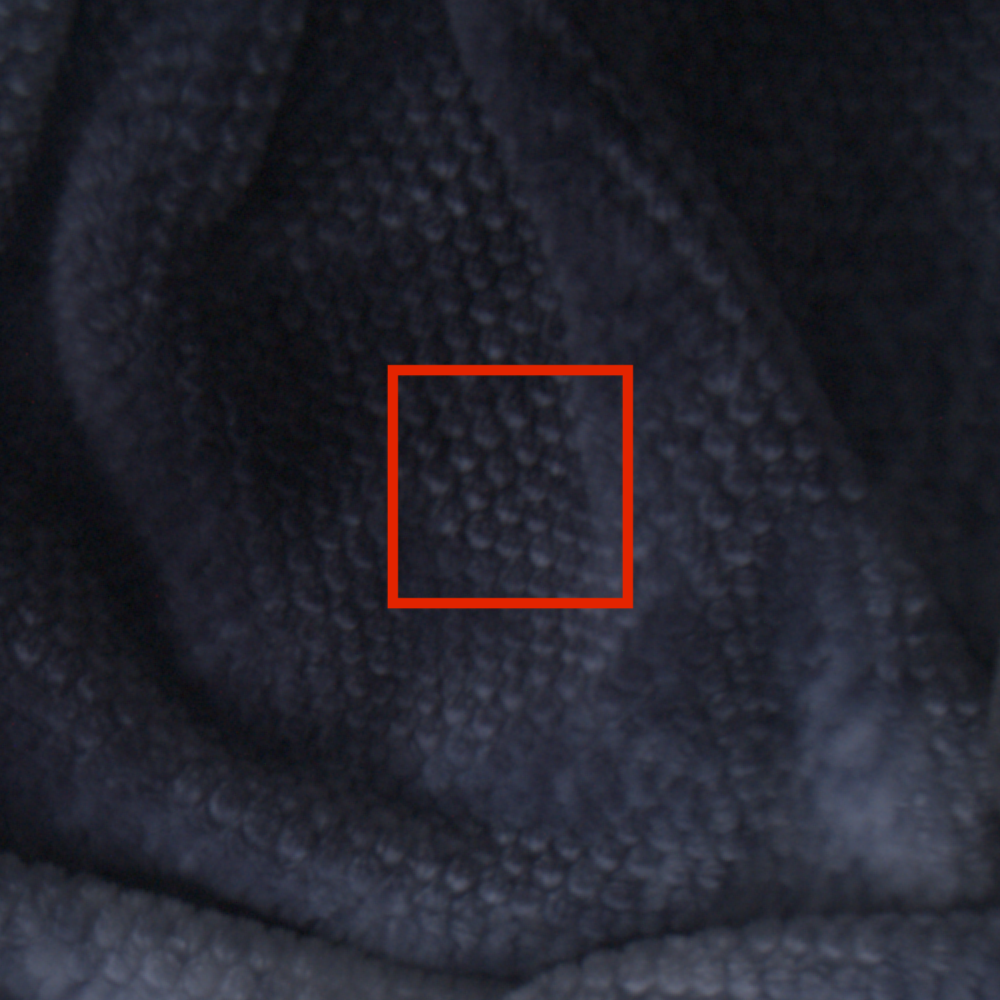}
    \adjustimage{width=1.0in,valign=m}{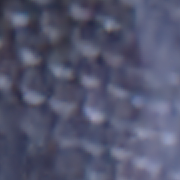} \adjustimage{width=1.0in,valign=m}{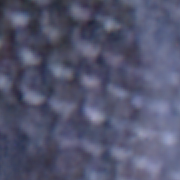}\\
    \caption{Summary of our improvements. Left: Source image. Red box highlights selected crop for comparison. Middle: Model output with wavelet transform, no NSMA, and Charbonnier Loss. Right: Model output with our Feature-Align layer, NSMA (n=4), and Feature Matching Loss. Middle and right images are brightened for clarity. Our method shows improved texture restoration.}
    \label{fig:summary1}
\end{figure}

\begin{figure}[!tb]
\centering
\begin{tabular}{c c}
 \rotatebox[origin=c]{90}{Noisy} &
 \adjustimage{width=1.0in,valign=m}{images/0_noisy_bright3.png}  \adjustimage{width=1.0in,valign=m}{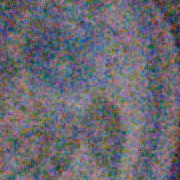}  \adjustimage{width=1.0in,valign=m}{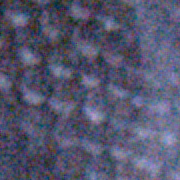} \\
 \rotatebox[origin=c]{90}{Feature-Align} &
 \adjustimage{width=1.0in,valign=m}{images/0_spade_bright3.png}  \adjustimage{width=1.0in,valign=m}{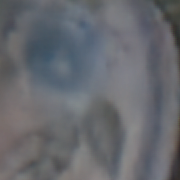}  \adjustimage{width=1.0in,valign=m}{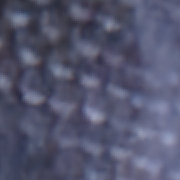} \\
 \rotatebox[origin=c]{90}{+ NSMA} &
 \adjustimage{width=1.0in,valign=m}{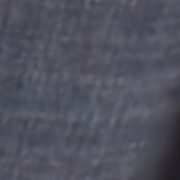}  \adjustimage{width=1.0in,valign=m}{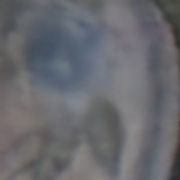}  \adjustimage{width=1.0in,valign=m}{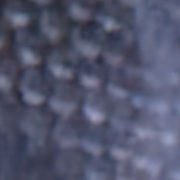} \\
 \rotatebox[origin=c]{90}{+ Feat. Match.} &
 \adjustimage{width=1.0in,valign=m}{images/0_feature_subrange_bright3.png}  \adjustimage{width=1.0in,valign=m}{images/3_feature_subrange_bright.png}  \adjustimage{width=1.0in,valign=m}{images/9_feature_subrange_bright.png} \\
 \rotatebox[origin=c]{90}{Ground Truth} &
 \adjustimage{width=1.0in,valign=m}{images/0_gt_bright3.png}  \adjustimage{width=1.0in,valign=m}{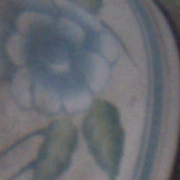}  \adjustimage{width=1.0in,valign=m}{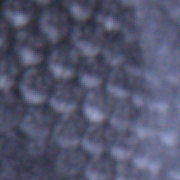} \\
 \end{tabular}
\caption{Our improvements in order: the effect of using the NSMA (n=4) on our Feature-Align method, followed by the effect of replacing Charbonnier Loss with Feature Matching Loss in the NSMA of Feature-Align models. Images are brightened for clarity.}
\label{fig:summary2}
\end{figure}

Figures \ref{fig:summary1} and \ref{fig:summary2} display a summary of our denoising improvements.

\clearpage

\begin{table*}[!tb]
\centering
\begin{tabular}{c || c c c c || c}
 \hline
 \multicolumn{6}{|c|}{Model Architecture Comparison} \\
 \hline
 Method & RAW PSNR & RAW SSIM & RGB PSNR & RGB SSIM & GMACs/MP\\
 \hline
 Input & 36.616 & 0.80509 & 27.581 & 0.51557 & - \\
 Plain U-Net & 45.485 & 0.97711 & 34.519 & 0.93481 & 2.2840 \\
 Wavelet & \textbf{45.576} & \textbf{0.97892} & \textbf{34.591} & 0.93157 & \textbf{1.9800} \\
 Feature-Align & 45.520 & 0.97780 & 34.565 & \textbf{0.93491} & 2.6688\\
 FA w/o connections & 45.395 & 0.97769 & 34.487 & 0.92534 & 2.5688 \\
 \hline
\end{tabular}

\begin{tabular}{c c c c c c}
Input & Plain U-Net & Wavelet & Feature-Align & FA w/o connections & Ground truth \\
\adjustimage{width=1.0in,valign=m}{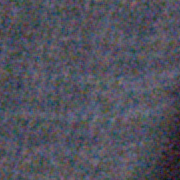} & \adjustimage{width=1.0in,valign=m}{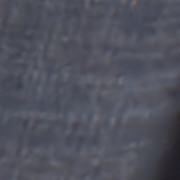} & \adjustimage{width=1.0in,valign=m}{images/0_wav_bright3.png} &
\adjustimage{width=1.0in,valign=m}{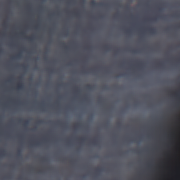} &
\adjustimage{width=1.0in,valign=m}{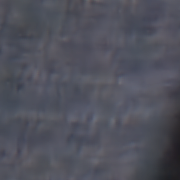} &
\adjustimage{width=1.0in,valign=m}{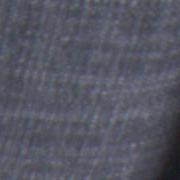}
\end{tabular}
\caption{Model with a Feature-Align layer shows a sharper image compared to a baseline wavelet transform model.}
\label{table:model_architecture}
\end{table*}

\begin{table*}[!tb]
\centering
\begin{tabular}{c || c c c c}
 \hline
 \multicolumn{5}{|c|}{Effect of Noise Subrange Model Array}\\
 \hline
 Method & RAW PSNR & RAW SSIM & RGB PSNR & RGB SSIM\\
 \hline
 Input & 36.616 & 0.80509 & 27.581 & 0.51557\\
 No NSMA & 45.520 & 0.97780 & 34.565 & 0.93491\\
 NSMA (n=4) & \textbf{45.668} & \textbf{0.97936} & \textbf{34.869} & \textbf{0.94014}\\
 \hline
\end{tabular}

\begin{tabular}{c c c c}
Input & No NSMA & NSMA (n=4) & Ground truth \\
\adjustimage{width=1.3in,valign=m}{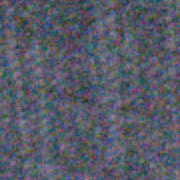} & 
\adjustimage{width=1.3in,valign=m}{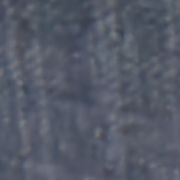} &
\adjustimage{width=1.3in,valign=m}{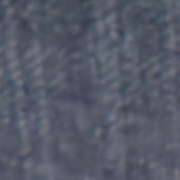} &
\adjustimage{width=1.3in,valign=m}{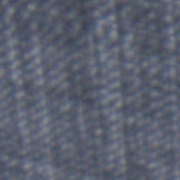}
\end{tabular}
\caption{NSMA substantially improves the image quality of our efficient model.}
\label{table:nsma}
\end{table*}

\begin{table*}[!tb]
\centering
\begin{tabular}{c || c c c c c}
 \hline
 \multicolumn{6}{|c|}{Effect of Loss Function} \\
 \hline
 Method & RAW PSNR & RAW SSIM & RGB PSNR & RGB SSIM & RGB LPIPS\\
 \hline
 Input & 36.616 & 0.80509 & 27.581 & 0.51557 & 0.53680\\
 Charbonnier & 45.668 & 0.97936 & 34.869 & 0.94014 & 0.28882\\
 RGB Perceptual & 45.519 & 0.97685 & 34.665 & 0.92168 & 0.32448 \\
 Knowledge Dist. & \textbf{45.813} & \textbf{0.98217} & \textbf{35.499} & \textbf{0.94079} & 0.27009 \\
 Feature Matching & 45.483 & 0.97828 & 34.607 & 0.93685 & \textbf{0.24603}\\
 \hline
\end{tabular}
\begin{tabular}{c c c c c c}
Input & Charbonnier & RGB Perceptual & Knowledge Dist. & Feature  & Ground truth \\
\adjustimage{width=1.0in,valign=m}{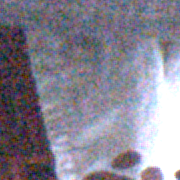} & 
\adjustimage{width=1.0in,valign=m}{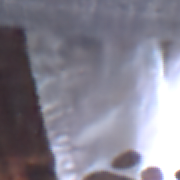} &
\adjustimage{width=1.0in,valign=m}{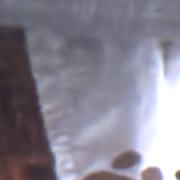} &
\adjustimage{width=1.0in,valign=m}{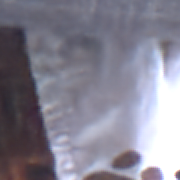} &
\adjustimage{width=1.0in,valign=m}{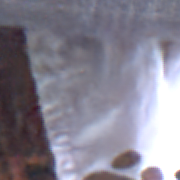} &
\adjustimage{width=1.0in,valign=m}{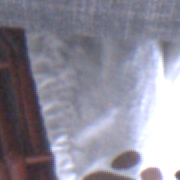}
\end{tabular}
\caption{Charbonnier Loss is outperformed by both Simple Knowledge Distillation and Feature Matching Loss. Simple Knowledge Distillation has the best performance in traditional metrics, while Feature Matching Loss has the best performance in the perceptual LPIPS metric. The model output with Feature Matching Loss appears to have the most detail.}
\label{table:loss}
\end{table*}

\clearpage

\subsection{Generalization}
For a standardized comparison of our method against existing state-of-the-art methods, we apply our method to the Darmstadt~Noise~Dataset~public~benchmark~\cite{darmstadt}. In training the noise parameters are tweaked to fit the DND test dataset and the white balance gains used to create the unprocessed MIRFLICKR dataset are tweaked to match ~\cite{brooks2019unprocessing} to be representative of DND. Table \ref{table:darmstadt} shows our method compared to the top 3 submissions on the benchmark. Our method achieves a RAW PSNR of 48.2756, a comparable metric to the CycleISP's~\cite{cycleisp} 49.1251 RAW PSNR, while using 263.32 times fewer MACs per megapixel. Our network uses 89,696 bytes of learned parameters, and the NSMA of 4 of our networks takes 4 times 89,696 bytes, which is 358,784 bytes. This is 22.7\% the size of learned parameters used by CycleISP, 1,578,624 bytes.

\begin{table}[!tb]
\vspace{-2.5mm}
\setlength{\tabcolsep}{7.3pt}
\scalebox{0.67}{
\centering
\begin{tabular}{c || c c c c || c }
 \hline
 \multicolumn{6}{|c|}{Darmstadt Noise Evaluation} \\
 \hline
 Method & RAW PSNR & RAW SSIM & RGB PSNR & RGB SSIM & GMACs/MP\\
 \hline
 Noisy & - & - & 29.836 & 0.7018 & - \\
 CycleISP & 49.1251 & 0.983 & 40.4987 & 0.9655 & 702.73  \\
 UPI Raw & 48.8905 & 0.9824 & 40.1728	& 0.9623 & 74.305 \\
 UPI RGB & 48.8824 & 0.9821 & 40.3545 & 0.9641 & 74.233  \\
 Ours & 48.2756 & 0.9808 & 39.5061 & 0.9572 & 2.6688  \\
 \hline
\end{tabular}
\vspace*{-1.6mm}
}
\caption{Our method achieves comparable results to the computationally expensive state-of-the-art methods.}
\label{table:darmstadt}
\end{table}

\section{Conclusion}
In this work we propose three innovations that enable high quality image denoising in an efficient model architecture: a Feature-Align layer, the use of an array of models tuned to different regions of the problem space, and a loss function for knowledge distillation that maximizes texture recovery. We combine these innovations along with existing state-of-the-art techniques and prove their efficacy by evaluating the method on a public benchmark. In addition, we propose a new public dataset of carefully constructed pairs of noisy and ground truth images with noise level annotations. The low computational cost of our method in comparison to existing state-of-the-art methods enables a variety of new applications for learning-based denoising in edge devices.

\clearpage

{\small
\bibliographystyle{ieee_fullname}
\bibliography{egbib}
}

\clearpage

\title{Feature-Align Network and Knowledge Distillation for Efficient Denoising \\ - Supplemental -}

\author{
\begin{tabular}{c@{\hspace{0.4in}}c@{\hspace{0.4in}}c}
Lucas D. Young$^*$& Fitsum A. Reda$^{*\dagger}$ & Rakesh Ranjan \\
Jon Morton &  Jun Hu &  Yazhu Ling \\
Xiaoyu Xiang & David Liu & Vikas Chandra 
\end{tabular} \\
Facebook Inc.
}

\maketitle
\ificcvfinal\thispagestyle{empty}\fi

\section{Noise Modeling}
\label{sec:noise}

Image noise originates at the Bayer RAW domain. The observed signal of a pixel is gaussianly distributed about the noiseless intensity of the pixel. Noise originates from a variety of sources, but can be broken down into two categories - signal independent noise and signal dependent noise. The variance of the gaussian distribution of signal independent noise per-pixel does not vary depending on the intensity of a pixel. The variance of signal dependent noise per-pixel is proportional to that pixel’s intensity. Read noise is a prominent example of signal independent noise and shot noise is a prominent example of signal dependent noise. 

\begin{equation}
\label{eq:gaussian}
y \sim \mathcal{N}(\mu=x, \sigma^2=ax + b)
\end{equation}

Equation \ref{eq:gaussian} models both signal independent noise and signal dependent noise as a single heteroscedastic gaussian, treating $y$ as a variable whose variance is a function of the true signal $x$, where $a$ and $b$ represent the signal dependent and signal independent noise respectively~\cite{azzari, practical, brooks2019unprocessing}. This description of noise is only partially complete because it does not account for clipping of the signal. Due to clipping, the per-pixel distribution of a pixel is a censored gaussian distribution, and the expected variance at the low and high ends of the signal are decreased compared to Equation \ref{eq:gaussian}~\cite{azzari}. To accurately model noise for a specific sensor, we choose realistic $a$ and $b$ parameters, sample the corresponding gaussian distribution for each pixel, add the noise to the ground truth image, and then clip the image.

\begin{figure*}[!htb]
    \centering
    \subfigure[Linear regression between gain and $a$]{\includegraphics[width=3.0in]{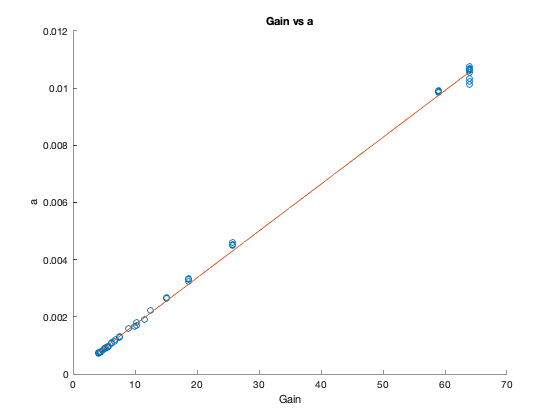}}
    \subfigure[Quadratic regression between gain and $b$]{\includegraphics[width=3.0in]{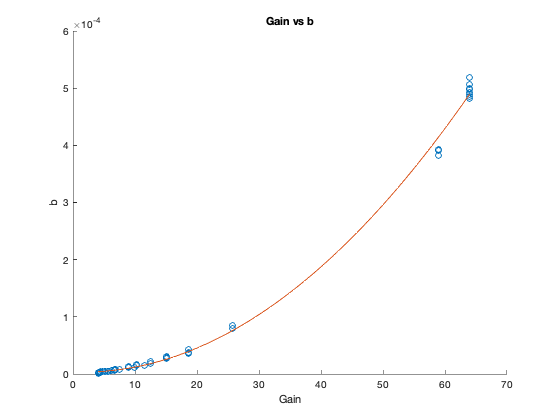}}
    \subfigure[Regression between $\log a$ and $\log b$]{\includegraphics[width=3.0in]{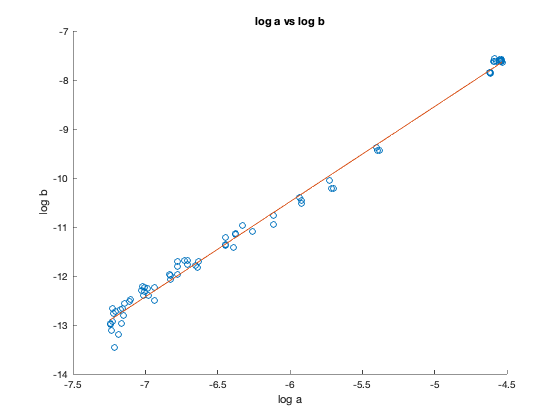}}
    \caption{The regressions in (a) and (b) are used to estimate noise parameters in inference. The regression in (c) is used to randomly choose noise parameters in training.}
    \label{fig:regressions}
\end{figure*}

To pick realistic $a$s and $b$s for the target sensor, we conduct an empirical analysis. In a lab-controlled lighting environment, we capture samples of frames of a color checker in a variety of lighting conditions, holding exposure time constant. We scale the image's intensities to a 0-1 range in accord with the camera's white and black levels. After segmenting color boxes in the checker, we approximate the noise-free intensity of each pixel as the mean intensity of its corresponding color box. This allows us to observe the relationship between noise-free intensity and variance for each image (see Figure \ref{fig:regressions}).

\begin{figure}
    \centering
    \includegraphics[width=2.0in]{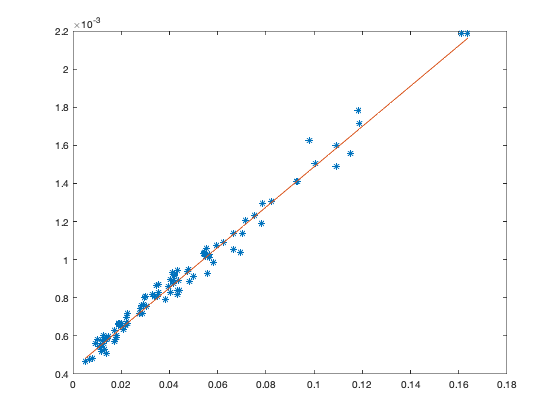}
    \caption{Example of intensity (x-axis) vs variance (y-axis) for a single image of a color checker.}
    \label{fig:variancevsintensity}
\end{figure}

We use the algorithm proposed by~\cite{azzari} to fit Equation \ref{eq:gaussian} to this plot. Note that while the images used for this analysis are clipped, the algorithm takes this discrepancy into account when fitting the theoretical unclipped noise model. This yields the parameters $a$ and $b$ for the image, which can be used to generate artificial noise to be subsequently added to a ground truth image and clipped.

In training, we model the distribution of these $a$ and $b$ pairs in logarithmic space, randomly choose $\log a$ along a uniform distribution, and then pick $\log b$ based on a linear regression. The resulting regression is unique to each type of sensor. Our noise modeling was done for target camera's sensor, the Sony IMX258. Since the image's gain is known at inference, to predict the noise levels at inference we create regressions between $a$ and gain and between $b$ and gain. We find that a linear regression fits the relationship between $a$ and gain and that a quadratic regression fits the relationship between $b$ and gain. As demonstrated in~\cite{practical}, superior denoising performance can be achieved by having this information available to the algorithm.

\section{Full Training and Implementation Details}
\label{sec:traindetails}
To train our array of models we use the following configuration:
\begin{itemize}
    \item The training ground truth images consist of unprocessed MIRFLICKR~\cite{mirflickr} with modified white balance gains for our target sensor, SIDD~\cite{sidd}, and the Learning~to~See~in~the~Dark~training~dataset~\cite{sid}. The latter two datasets are unified into an RGGB pattern with Bayer~Unification~\cite{bayeraug}. These images are randomly cropped into 128 x 128 patches.
    \item Bayer~Augmentation~\cite{bayeraug} is applied to the training data.
    \item The input to the model is generated by adding artificial noise to the ground truth image. A description of our experiments to realistically model artificial noise is included in supplemental Section \ref{sec:noise}.
    \item The k-sigma transform of~\cite{practical} is implemented in training and testing. Note that $k$ and $\sigma^2$ refer to our $a$ and $b$ noise parameters respectively.
    \item Training examples are collated into batches of size 16.
    \item The Charbonnier Loss variant of our models uses a loss weight of 393.5. The Feature Matching Loss variant of our models uses a loss weight of 78.7. These weights were derived from a hyperparameter sweep on RAW PSNR.
    \item Models are trained using the Adam~\cite{adam} optimizer with the maximum learning rate of 1e-4.
    \item Training occurs in 2,500,000 iterations scheduled with cosine learning rate decay.
\end{itemize}

\section{Test Dataset Details}
\label{sec:testdetails}
We collect noisy-clean image pairs on the targeted camera of our method which uses a Sony IMX258 sensor. Pairs are collected by taking a short exposure with a random gain between 1.0 and 64.0 accompanied by a long exposure of the same scene with a gain of 1.0. The long exposure's exposure time is adjusted such that the brightness of both images are equivalent. The gain and exposure time is selected programatically so that the camera is not moved slightly between image captures. Similar to the Darmstadt Noise Dataset ~\cite{darmstadt}, to account for small environmental vibrations occurring between the short and long exposure that misalign the pair, we predict and correct for a global 2D translation estimated by averaging the Lucas-Kanade~\cite{lucaskanade} optical flow of features detected by the Shi-Tomasi ~\cite{goodfeatures} algorithm. Finally, noise level annotations $a$ and $b$ for the short exposure are estimated from the regression described in supplemental Section \ref{sec:noise}.
\end{document}